\documentclass[aps,prd,twocolumn,groupedaddress,nofootinbib,showpacs,eqsecnum]{revtex4}

\usepackage{euscript,graphicx,amssymb,subfigure}

\usepackage{amsfonts}
\usepackage{amsmath}
\usepackage{amssymb}
\usepackage{graphicx}
\usepackage{euscript}
\usepackage{color}
\usepackage{subfigure}

\newcommand{\goodgap}{\hspace{\subfigtopskip} \hspace{\subfigbottomskip}}

\begin{document}

\title{Cosmography of $f(R)$\,-\,brane cosmology}

\author{Mariam Bouhmadi-L\'{o}pez$^1$\footnote{{\tt mariam.bouhmadi@ist.utl.pt}}, Salvatore Capozziello$^{2,3}$\footnote{{\tt capozziello@na.infn.it}}, Vincenzo F. Cardone$^{2,4}$\footnote{{\tt winnyenodrac@gmail.com}}}

\affiliation{$^1$ Centro Multidisciplinar de Astrof\'{\i}sica - CENTRA, Departamento de F\'{\i}sica, Instituto Superior T\'ecnico, Av. Rovisco Pais 1, 1049-001 Lisboa, Portugal}

\affiliation{$^2$ Dipartimento di Scienze Fisiche, Universit\`{a} di Napoli ``Federico II'', Compl. Univ. Monte S. Angelo, Ed.N, Via Cinthia, I-80126 Napoli, Italy}

\affiliation{$^3$ I.N.F.N. - Sez. di Napoli, Compl. Univ. Monte S. Angelo, Ed.G, Via Cinthia, I-80126 Napoli, Italy}

\affiliation{$^4$ Dipartimento di Scienze e Tecnologie dell' Ambiente e del Territorio, Universit\`{a} degli Studi del Molise, Contrada Fonte Lappone, 86090\,-\,Pesche (IS), Italy}


\begin{abstract}
Cosmography is a  useful tool  to constrain cosmological models,
in particular dark energy models.  In the case of  modified
theories of gravity, where the equations of motion are generally
quite complicated, cosmography can contribute to select realistic
models without imposing arbitrary choices {\it a priori}. Indeed,
its reliability  is based on the assumptions that the universe is
homogeneous and isotropic on large scale and luminosity distance
can be "tracked" by the derivative series of the scale factor
$a(t)$. We apply this approach to induced gravity brane-world
models where  an $f(R)$-term is present in the brane effective
action. The virtue of the model is to self-accelerate the normal
and healthy DGP branch once the $f(R)$-term deviates from the
Hilbert-Einstein action. We show that the model, coming from a
fundamental theory, is consistent with the $\Lambda$CDM scenario
at low redshift. We finally estimate the cosmographic parameters
fitting the Union2 Type Ia Supernovae (SNeIa) dataset and the distance
priors from Baryon Acoustic Oscillations (BAO) and then provide constraints
on the present day values of $f(R)$ and its second and third derivatives.
\end{abstract}

\pacs{98.80.-k,98.80.Es,11.10.-z}

\maketitle

\section{Introduction}

The late-time acceleration  of the Universe has been confirmed by
several observations ranging from type Ia Supernovae (SNeIa)
\cite{Perlmutter:1998np}, which brought the first evidence, to the
cosmic microwave background (CMB) \cite{Spergel:2003cb} and the
baryon acoustic oscillations (BAO) \cite{Cole:2005sx}. More
recently,  gamma ray bursts (GRB), also if not properly standard
candles,  have been as well very useful at this regard
\cite{izzo,Cardone:2010rr}. They could, in principle, be useful to
probe high redshifts with the aim to remove degeneracy of
cosmological models with respect to $\Lambda$CDM
\cite{izzo2,izzo3}. While the recent speed up of the universe is a
fact, we have yet no answer to the question: What is the ``hand
that rocks the cradle''?

If we  assume that general relativity is valid on all the scales,
even though it has been corroborated at most on the solar system
range, then we require a component on the budget of the universe,
that violates at least the strong energy condition to describe the
current acceleration of the universe \cite{Copeland:2006wr}. The
simplest option at this regard corresponds to a cosmological
constant, giving raise to the $\Lambda$CDM model which matches
pretty well the observations, but then  we face the cosmological
constant problem. An alternative approach is to invoke a
gravitational theory that deviates from general relativity on the
appropriate scales and at the same time being able to reproduce
the big achievements of general relativity (cf.
Refs.~\cite{Nojiri:2006ri,Capozziello:2007ec,Sotiriou:2008rp,Durrer:2008in,DeFelice:2010aj}).
The latter approach can be tackled in the context of brane-world
models \cite{Maartens:2010ar}, which are inspired in string
theory, where our universe corresponds to a 4-dimensional
hypersurface embedded on the higher dimensional space-time,
usually dubbed the bulk. Several approach have been undertaken,
for example in the context of induced gravity brane-world
\cite{brane,BouhmadiLopez:2004ys} the self-accelerating brane of
the Dvali-Gabadadze-Porrati (DGP) model is probably the most
famous \cite{dgp}.

The DGP model has gathered a  lot of attention on the last years.
As an induced gravity brane-world model, it contains two possible
solutions, the self-accelerating branch, which is asymptotically
de Sitter, and the normal branch. Despite this fact, the
self-accelerating brane does not require any type of dark energy
to describe a late-time inflationary  period of the brane, it
suffers from some theoretical problems like the ghost problem
\cite{Koyama:2007za}; i.e. a degree of freedom that shows up when
the brane is perturbed and behaves on the brane effectively as a
scalar field with the wrong kinetic energy. On the other hand, the
normal branch is ``healthy'' in the sense that it does not  suffer
from the ghost problem but it requires some sort of dark energy to
describe the late-time acceleration of the universe.

In a previous paper \cite{BouhmadiLopez:2009db},  one of us
proposed a mechanism to self-accelerate the normal DGP branch.
More precisely,  a generalized induced gravity brane-world model
is proposed where the brane action contains an arbitrary $f(R)$
term, $R$ being the scalar curvature of the brane\footnote{See
Ref.~\cite{Balcerzak:2010kr} for a brane-world model with an
$f(R)$ term in the bulk.}. It is shown that an $f(R)$ ($\neq R$)
term on the dynamics of a homogeneous and isotropic brane induces
a shift on the energy density of the brane. This new shift term,
which  is absent in the DGP model, plays a crucial role to
self-accelerate the generalized normal DGP branch of the model. In
other terms, the generalized normal branch is asymptotically de
Sitter without considering any dark energy on the brane.

In the present paper, we discuss the possibility to constrain this
model using a cosmographic approach \cite{Capozziello:2008qc}.
Cosmography relies  on two crucial things: i) extracting the
maximum amount of information from measured distances, like the
luminosity distances of SNeIa, ii) assuming  that the universe can
be modelled by a Friedmann-Lema\^{\i}tre-Robertson-Walker (FLRW)
model on large scale without assuming {\it a priori} any dynamical
theory to describe it. Now, why have we chosen this approach? for
several reasons: i) for its simplicity. For example, the modified
Einstein equation of the brane are of fourth order on the scale
factor (due to the $f(R)$-term in the brane action) and therefore
very difficult to solve analytically. In the cosmographic approach
we do not need to have an explicit solution for the evolution of
the scale factor in terms of the cosmic time of the brane. ii) The
approach is quite general in the sense that we do not have to
specify which $f(R)$ function we are dealing with. The only
requirement is that $f(R)$ is an analytic function.

The outline of the paper  is as follows. In Sect.II, we review the
model presented in \cite{BouhmadiLopez:2009db}. In particular, we
highlight how the model contains fixed points corresponding to de
Sitter solutions (in absence of any matter on the brane); i.e.
self-accelerating solutions. In Sect.III, we present the
cosmographic approach we will follow. We write down all the
quantities relevant of the model in terms of the cosmographic
parameters. In Sect.IV, we constrain the model from a theoretical
point of view, while Sect.V deals with observational constraints. 
Finally, we summarize and discuss the obtained results in Sect.VI.

\section{An $f(R)$-term on the brane}

In this section, we review the model  introduced in
\cite{BouhmadiLopez:2009db}. The scenario corresponds to a
5-dimensional  brane-world model whose action reads
\begin{eqnarray}
\mathcal{S} &=& \,\,\, \int_{\mathcal{B}} d^5X\, \sqrt{-g^{(5)}}\;
\left\{\frac{1}{2\kappa_5^2}R[g^{(5)}]\;\right\}\nonumber \\
&& + \int_{h} d^4X\, \sqrt{-g}\; \left\{\frac{1}{\kappa_5^2} K\;+\;\frac{1}{2\kappa_4^2} f(R) + \mathcal{L}_m \right\}\,, \label{action}
\end{eqnarray}
where $\kappa_5^2$ is the 5D gravitational  constant, $R[g^{(5)}]$
is the scalar curvature in the bulk and $K$ the extrinsic
curvature of the brane in the higher dimensional bulk. For the
sake of simplicity, we have assumed a vanishing bulk cosmological
constant, for a more general setup please see
\cite{BouhmadiLopez:2009db}. In addition, $R$ is the scalar
curvature of the induced metric on the brane, $g$, and
$\kappa_4^2$ is related to the Newtonian gravitational constant,
$G$, through $\kappa_4^2=8\pi G$. The function $f(R)$ has mass
square units. On the other hand, $\mathcal{L}_m$ corresponds to
the standard matter Lagrangian of the brane. We recover the DGP
model \cite{dgp,brane} when $f(R)=R$.

From now on, we assume a homogeneous and isotropic  brane with
spatially flat sections. Therefore, the modified Friedmann
equation can be written as
\begin{eqnarray}
3H^2= \frac{\kappa_5^4}{12}\rho^2.\label{friedmann}
\end{eqnarray}
The  total energy density $\rho$ is conserved and is given by
\begin{equation}
\rho=\rho_m+\rho_f,
\end{equation}
where
\begin{eqnarray}
\rho_m&=& \frac{\rho_{m0}}{a^3},\nonumber \\
\rho_f&=& -\frac{1}{\kappa_4^2}\left[3 H^2 f'-\frac12(Rf'-f)+3H\dot R f'' \right], \nonumber \\
\end{eqnarray}
where both energy densities $\rho_m$ and $\rho_f$  are conserved
separately. We will use the  subscript $0$ to refer to quantities
evaluated at the present time. The dot stands for derivative with
respect to the cosmic time of the brane and the prime for
derivative respect to the scalar curvature of the brane.

We are interested on the branch that generalize the  standard DGP
solution and therefore the modified Friedmann equation
(\ref{friedmann}) reduces to
\begin{eqnarray}
H= \frac{\kappa_5^2}{6}\rho.\label{friedmann2}
\end{eqnarray}
The other root of Eq.~(\ref{friedmann}) generalizes the Friedmann
equation of the self-accelerating DGP solution.

For latter convenience it is useful to rewrite Eq.~(\ref{friedmann2}) as
\begin{equation}
f'H^2+\frac{1}{r_c}H=\frac{\kappa_4^2}{3}\frac{\rho_{m0}}{a^3} + \frac16\left(Rf'-f-6H\dot R f''\right).
\label{friedmann3}
\end{equation}
The parameter $r_c = \kappa_5^2/(2\kappa_4^2)$ is the  crossover
scale. For $1\ll f'r_c H$, we obtain the Friedmann equation for
4-dimensional $f(R)$ models.

The Raychaudhuri equation for this model can be deduced  by taking
the time derivative of Eq.~(\ref{friedmann3}), bearing in mind
that the matter energy density is conserved, and it reads
\begin{eqnarray}
\dot H +\frac {1}{r_c} \frac{1}{2f'}\frac{\dot H}{H}= -\frac{\kappa_4^2}{2f'} \frac{\rho_{m0}}{a^{3}}-\frac{\dot R^2 f'''+(\ddot R- H \dot R)f''}{2f'}. \nonumber \\
\label{raychaudhuri}
\end{eqnarray}
To obtain this equation we have as well  used\footnote{We use
Wald's book sign convention.} $R=6(2H^2+\dot H)$.

It can be shown that the brane contains  fixed points
corresponding to de Sitter solutions (once the matter content is
negligible) \cite{BouhmadiLopez:2009db}, therefore the brane
enters a self-accelerating regime at some point along its
expansion. In reference \cite{BouhmadiLopez:2009db}, it is  shown
 what are the conditions to be fulfilled for the de Sitter
solutions to be stable under  homogeneous perturbations
\cite{BouhmadiLopez:2009db}. More precisely, we can associate an
effective square mass to the perturbations and, as long as this
quantity is positive, we can conclude that de Sitter solution is
stable.

\section{Cosmography}

\subsection{General approach}

As we said, cosmography relies on the assumption that the universe
is homogeneous and isotropic on large scale and   no dynamical
theory  is assumed a priori \cite{Capozziello:2008qc}. In
particular it relies on the scale factor series expansion  of a
FLRW metric in terms of time \cite{Capozziello:2008qc}; i.e.
\begin{eqnarray}
\frac{a(t)}{a(t_0)}&=&1+H_0(t-t_0)-\frac{q_0}{2}H_0^2(t-t_0)^2\nonumber\\&\,&+ \frac{j_0}{3!}H_0^3(t-t_0)^3+\frac{s_0}{4!}H_0^4(t-t_0)^4\nonumber \\
&\,&+\frac{l_0}{5!}H_0^5(t-t_0)^5 + O((t-t_0)^6)
\label{expansionat}
\end{eqnarray}
where the standard cosmographic parameters are defined as \cite{Capozziello:2008qc}
\begin{equation}
\begin{array}{l}
\displaystyle{H = \frac{1}{a} \frac{da}{dt}} \\ ~ \\ \displaystyle{q = - \frac{1}{a} \frac{d^2a}{dt^2} \ H^{-2}}
\\ ~ \\ \displaystyle{j = \frac{1}{a} \frac{d^3a}{dt^3} \ H^{-3}} \\ ~ \\
\displaystyle{s = \frac{1}{a} \frac{d^4a}{dt^4} \ H^{-4}} \\ ~ \\
\displaystyle{l = \frac{1}{a} \frac{d^5a}{dt^5} \ H^{-5}}.
\end{array}
\label{cosmopar}
\end{equation}

These parameters are usually referred to as the Hubble,
deceleration, jerk, snap and lerk parameters respectively (see
\cite{Capozziello:2008qc} and references therein). Their present
day values (which we will denote with a subscript $0$) can be used
to characterize the evolutionary status of the Universe. For
example, $q_0 < 0$ denotes an accelerated expansion, while a
change of sign of $j$ (in an expanding universe) signals that the
acceleration starts increasing or decreasing.

Most importantly, the parameters $\{q_0,j_0,l_0,s_0\}$ can be
used to evaluate different distances in the universe. This can be
achieved by inverting the relation (\ref{expansionat}) and bearing
in mind that the distance, $D$, travelled by a given photon that
was emitted at $t_1$ and detected at the current epoch $t_0$ is
simply $D=t_0-t_1$ (where we have set the speed of light to
unity). Therefore, one can obtain a series expansion of the
distance $D$ in terms of the scale factor or redshisft, while the
coefficients of the expansion are defined through the cosmographic
parameters \cite{Capozziello:2008qc}. The distance $D$ can be
related to several physical magnitude, for example the luminosity
distance, the angular diameter distance and many more
\cite{Cattoen:2008th}. These magnitudes can be constrained
observationally through SNeIa, BAO and, possibly, GRB data
\cite{izzo}. In fact, these data are useful to construct a cosmic
ladder where any step is a cosmic indicator. Once the distances
are constrained, we obtain as well constraints on the values
acquired by the cosmographic parameter (see for example
\cite{Capozziello:2008qc,Cattoen:2008th,Vitagliano:2009et}). It is
worthy to notice, at this regard, that given that the cosmographic
approach is based on a Taylor expansion of the scale factor, or
redshift, for data of GRB at high redshift (above $z=1$), it is
better to use the variable  $y=z/(1+z)$, introduced in
\cite{polarsky}, instead of the redshift.

\subsection{Applying cosmography to  $f(R)$ brane-world}

In this subsection,  we will relate the characteristic quantities
 defining the model introduced in Sect.II to the parameters
$\{q_0,j_0,l_0,s_0\}$. In addition, this will be done without
specifying a particular $f(R)$ model on the brane.

We start reminding that the derivative of the Hubble parameter can
be expressed in terms of the cosmographic parameters. Indeed,
after some algebra, the following relation can be obtained:

\begin{equation}
\dot{H} = -H^2 (1 + q) \ ,
\label{hdot}
\end{equation}

\begin{equation}
\ddot{H} = H^3 (j + 3q + 2) \ ,
\label{h2dot}
\end{equation}

\begin{equation}
\dddot{H} = H^4 \left [ s - 4j - 3q (q + 4) - 6 \right ] \ ,
\label{h3dot}
\end{equation}

\begin{equation}
d^4H/dt^4 = H^5 \left [ l - 5s + 10 (q + 2) j + 30 (q + 2) q + 24 \right ]
\ .
\label{h4dot}
\end{equation}

Now, the question is how our model can  be characterized  by these
parameters, or, more precisely, what can be said about the current
values of $f(R_0), f'(R_0), f''(R_0), f'''(R_0)$. In order to
answer this question we have first to rewrite $R, \dot R, \ddot R,
\dddot R$ in terms of $q,j,s,l$. This can be done with some
algebra

\begin{equation}
\begin{array}{l}
\displaystyle{\dot{R} = 6 \left ( \ddot{H} + 4 H \dot{H} \right )}, \\ ~ \\
\displaystyle{\ddot{R} = 6 \left ( \dddot{H} + 4 H \ddot{H} + 4 \dot{H}^2 \right )}, \\  ~ \\
\displaystyle{\dddot{R} = 6 \left ( d^4H/dt^4 + 4 H \dddot{H} + 12 \dot{H} \ddot{H} \right )}. \\
\end{array}
\ .
\label{prederr}
\end{equation}

Now using Eqs.~(\ref{hdot}), (\ref{h2dot}), (\ref{h3dot}) and
(\ref{h4dot}), we get

\begin{equation}
R = 6 H^2 (1 - q) \ ,
\label{rz}
\end{equation}

\begin{equation}
\dot{R}= 6 H^3 (j - q - 2) \ ,
\label{rdotz}
\end{equation}

\begin{equation}
\ddot{R}= 6 H^4 \left ( s + q^2 + 8 q + 6 \right ) \ ,
\label{r2dotz}
\end{equation}

\begin{equation}
\dddot{R} = 6 H^5 \left [ l- s - 2 (q + 4) j - 6 (3q + 8) q - 24 \right ] \ .
\label{r3dotz}
\end{equation}

If we substitute Eqs, (\ref{hdot}), (\ref{h2dot}), (\ref{h3dot}),
(\ref{h4dot}), (\ref{rz}), (\ref{rdotz}),  (\ref{r2dotz}) and
(\ref{r3dotz}) in the Friedmann and Raychaudhuri equations, and
evaluate them at the present time, we could obtain, in principle,
the current values of $f(R_0), f'(R_0), f''(R_0), f'''(R_0)$.
However as we have only two equations, the Friedmann relation and
the Raychaudhuri equation, we require more information to define
completely the model. At this respect, notice that the effective
gravitational constant on the brane $G_{\rm{eff}}=G/f'$ (see the
Friedmann equation (\ref{friedmann3})), therefore we can assume,
as a prior,  that $f'(R_0)=1$ such that the current value of the
gravitational constant coincides with the Newtonian one. Further
information can be obtained through the equation satisfied by
$\dddot H$. At this respect, we take the time derivative of
Eq.~(\ref{raychaudhuri}) and we obtain

\begin{eqnarray}
\ddot H &+& \frac{1}{2r_c} \frac{(\ddot H H- \dot H^2)f'-H\dot H f'' \dot R}{(Hf')^2}
=\nonumber \\
&&\frac{\dot R ^2f'''+(\ddot R-\dot R H)f''+\kappa_4^2\rho_{m0} a^{-3}}{2f'^2(f'' \dot R)^{-1}} \nonumber \\
&-&\frac{\dot R^3 f^{(iv)}+(3\ddot R \dot R -H\dot R^2)f'''}{2f'}\nonumber\\
&-&\frac{(\dddot R-\ddot R H-\dot R \dot H)f''-3\kappa_4^2 H\rho_{m0}a^{-3}}{2f'},
\label{dddotH}
\end{eqnarray}
where $f^{(iv)}=d^4f/dR^4$. As third assumption, we take into
account the power series
\begin{eqnarray}
f(R)&\simeq& f(R_0)+f'(R_0)(R-R_0)+\frac12 f''(R_0)(R-R_0)^2 \nonumber\\
&\,& +\frac16 f'''(R_0) (R-R_0)^3.\label{seriefr}
\end{eqnarray}
i.e. at low redshift, the $f(R)$-function is well approximated by
its Taylor expansion up to the third order\footnote{In what
follows we assume $f'(R_0)=1$ and $f^{(iv)}(R_0)\simeq 0$.}.

Now we can finally substitute Eqs. (\ref{hdot}), (\ref{h2dot}),
(\ref{h3dot}), (\ref{rz}), (\ref{rdotz}),  (\ref{r2dotz}) and
(\ref{r3dotz}) in the Friedmann constraint (\ref{friedmann3}), the
Raychaudhuri relation (\ref{raychaudhuri}) and the complementary
equation (\ref{dddotH}). We evaluate them at the present time.
Notice that Eqs.~(\ref{friedmann3}) and (\ref{raychaudhuri}) can
be expressed as linear combinations of $f(R_0), f'(R_0), f''(R_0),
f'''(R_0)$ at $z=0$. This is not the case for equation
(\ref{dddotH}) as it is quadratic on $f''(R_0$). So, we will
proceed as follows, we obtain $f(R_0)$ as a linear combination of
$f''(R_0)$ using Eq.~(\ref{friedmann3}),
\begin{equation}
f(R_0)=6\left[\left(\Omega_m-1\right)H_0^2+\frac16\left(R_0-6H_0\dot{R_0}f''\right)-\frac{1}{r_c}H_0\right],
\label{f}
\end{equation}
where $\Omega_0=\kappa_4^2\rho_{m_0}/(3H_0^2)$. Similarly, we can write $f'''(R_0)$
as a linear combination of $f''(R_0)$ using Eq.~(\ref{raychaudhuri}), i.e.
\begin{eqnarray}
f'''(R_0)=-\frac{3H_0^2\Omega_m+\dot{H_0}(2+\frac{1}{r_c H_0})
+ (\ddot{R_0}-H\dot{R_0})f''(R_0)}{\dot R_0^2}. \nonumber\\
\label{f3}
\end{eqnarray}
Then we rewrite Eq.(\ref{dddotH}) as follows
\begin{equation}
a_2f''(R_0)^2+a_1f''(R_0)+a_0=0,
\label{f2}
\end{equation}
where
\begin{eqnarray}
a_2&=&\dot R_0(\ddot R_0-\dot R_0 H_0),\\
a_1&=&\dot R_0^3 f'''(R_0)+3H_0^2\Omega_m\dot R_0-\left(\dddot R_0-\ddot R_0-\dot R_0 \dot H_0\right)\nonumber\\ & & +\frac{\dot H_0\dot R_0}{r_c H_0}\\
a_0&=&-\left(3\ddot R_0\dot R_0 -H_0 \dot R_0^2\right)f'''(R_0)+9H_0^3\Omega_m-2\ddot H_0 \nonumber\\
&-&\frac{\ddot H_0 H_0 -\dot H_0^2}{r_cH_0^2}.
\end{eqnarray}

Even though the previous equation  looks quadratic in $f''(R_0)$,
it is not the case because $a_1$ is a linear function of
$f'''(R_0)$ and therefore this term contributes quadratically in
$f''(R_0)$. Once we substitute Eq.~(\ref{f3}) on Eq.~(\ref{f2}),
we obtain a linear equation for $f''(R_0)$.

Finally, we obtain the following results:
\begin{eqnarray}
\frac{f(R_0)}{6H_0^2}&=&-\frac{\mathcal{A}_0\Omega_m+\mathcal{B}_0+\mathcal{C}_0(r_cH_0)^{-1}}{\mathcal{D}}, \\
\frac{f''(R_0)}{(6H_0^2)^{-1}}&=&-\frac{\mathcal{A}_2\Omega_m+\mathcal{B}_2+\mathcal{C}_2(r_cH_0)^{-1}}{\mathcal{D}}, \\
\frac{f'''(R_0)}{(6H_0^2)^{-2}}&=&-\frac{\mathcal{A}_3\Omega_m+\mathcal{B}_3+\mathcal{C}_3(r_cH_0)^{-1}}{(j_0-q_0-2)\mathcal{D}},
\label{eq: frfinal}
\end{eqnarray}
where $\mathcal{A}_i,\mathcal{B}_i,\mathcal{C}_i$ and
$\mathcal{D}$ with $i=0,2,3$ are functions of $q,j,s,l$ which are
defined as

\begin{eqnarray}
\mathcal{A}_0&=&\left( j_0-q_0-2 \right)l_0\nonumber\\
&-&\left(3s_0+7j_0+ 6q_0^{2}+41q_0+22\right)s_0\nonumber\\
&-&\left[\left(3q_0+16\right)j_0+20q_0^{2}+64q_0+12\right]j_0\nonumber \\
&-&\left(3q_0^{4}+25q_0^{3}+96q_0^{2}+72q_0+20\right),\\
\mathcal{B}_0&=&-\left(j_0q_0-q_0^{2}-2q_0\right)l_0\nonumber\\
&+&\left[3q_0s_0+\left(4q_0+6\right)j_0+6q_0^{3}+44q_0^{2}+22q_0 \right.\nonumber\\
&\,&\left.-12\right]s_0\nonumber\\
&+&\,\left[2j_0^2+ \left(3q_0^{2}+10q_0-6 \right) j_0+ 17q_0^3+52q_0^{2}\right.\nonumber\\
&\,&\left.+54q_0 +36\right] j_0\nonumber\\
&+&3q_0^{5}+28q_0^{4}+118q_0^{3}+72q_0^{2}-76q_0-64,\\
\mathcal{C}_0&=& -\left(j_0-q_0-2 \right) l_0\nonumber\\
&+&\left[3s_0+ \left(3q_0+1\right)j_0+3\,q_0^2+41q_0 +34\right]s_0\nonumber\\
&+&\left[j_0^2-\left(q_0^2-q_0-6\right)j_0+5q_0^3+43q_0^{2}\right.\nonumber\\
&\,&\left.+50q_0+4\right]j_0\nonumber\\
&-&(q_0^4+3q_0^3-80q_0^2-144q_0-68),
\label{eq: a0b0c0}
\end{eqnarray}
\begin{eqnarray}
\mathcal{A}_2&=& 9s_0+6j_0+9q_0^2+66q_0+42,\\
\mathcal{B}_2&=& -\left\{6\left(q_0+1\right) s_0+2\left[j_0+\left(q_0-1\right)\right]j_0\right.\nonumber\\
&+&\left.6q_0^3+50q_0^2+74q_0+32\right\},\\
 \mathcal{C}_2&=&-\left\{3\left(1+q_0\right) s_0+\left[j_0-(q_0^2+q_0+2)\right]j_0\right.\nonumber\\
&+&\left. 4q_0^3+29q_0^2+42q_0+18\right\},
\label{eq: a2b2c2}
\end{eqnarray}
\begin{eqnarray}
\mathcal{A}_3&=& -3\left[l_0+s_0-3\left(q_0+4\right)j_0-15q_0^2\right.\nonumber\\
&\,&\left.-26q_0-4\right],\\
\mathcal{B}_3&=&2\left[\left(1+q_0\right)l_0+\left(q_0+j_0\right)s_0\right.\nonumber\\
&\,&\left.-\left(j_0+2q_0^2+6q_0+3\right)j_0\right.\nonumber\\
&\,&-\left. \left(15q_0^3+42q_0^2+39q_0+12\right)\right],\\
\mathcal{C}_3&=&\left(1+q_0\right)l_0+\left(j_0-q_0^2-q_0-1\right)s_0 \nonumber\\
&-&\left(j_0+q_0^2+4q_0+2\right)j_0 \nonumber\\
&-&(q_0^4+26q_0^3+69q_0^2+64q_0+20),\\
%
\mathcal{D}&=& -\left(j_0-q_0-2\right)l +\left(3s_0-2j_0+6q_0^2+50q_0+40\right)s_0\nonumber\\
&+&\left[\left(3q_0+10\right)j+11q_0^2+4q_0-18\right]j_0\nonumber\\
&+&3q_0^4+34q_0^3+180q_0^2+246q_0+104.
\label{eq: a3b3c3}
\end{eqnarray}

We have split  the expressions of $f(R_0)$, $f''(R_0)$ and
$f'''(R_0)$  into three pieces involving the functions
$\mathcal{A}_i\Omega_m$, $\mathcal{B}_i$ and
$\mathcal{C}_i(r_cH_0)^{-1}$, where $\mathcal{A}_i$,
$\mathcal{B}_i$ and $\mathcal{C}_i$  are defined exclusively in
terms of the cosmographic parameters. The first term
$\mathcal{A}_i\Omega_m$ account for the contribution of matter to
the $f(R)$-function\footnote{It is worth noticing that we are
developing our considerations in the Jordan frame so the standard
matter is minimally coupled to the geometry.}. The second one
$\mathcal{B}_i$ is a purely geometrical one. The third one takes
into account the effect of the extra dimension; i.e. it involves
the crossover scale $r_c$. Not surprisingly,  if we switch off
this term; i.e. $1 \ll r_c$, we recover exactly the results
obtained in \cite{Capozziello:2008qc} corresponding to a standard
4-dimensional $f(R)$ scenario.

In summary, for a given set of values of the cosmographic
parameters we can deduce the function $f(R)$ through the
expression (\ref{seriefr}). Notice that the opposite is not
possible because the equations (3.23)-(3.32) are non-linear in
$\{q_0,s_0,l_0,j_0\}$. Moreover, by specifying a given function
$f(R)$, we do not obtain a unique evolution for the brane because
the modified Raychaudhuri equation is of fourth order in the scale
factor.

\section{Parameterizing the cosmographic parameters}

In order to get a first hint on the possible values of $f(R)$ and its derivatives
we adopt the following strategy: the cosmographic parameters will be calculated
for a given dark energy phenomenological parameterization. The
best and simplest one is the $\Lambda$CDM model. Next, we will
evaluate those parameters using the recent data of WMAP7 and the
constraint on the crossover scale $r_c$ (see \cite{Lazkoz:2007zk}
for details). Through  these results, we can constrain the $f(R)$
function as we will show below. This is a minimal approach but it
is useful to probe the self-consistency of the model.

The cosmographic parameters for the $\Lambda$CDM model read
\begin{eqnarray}
q&=&-\left(\frac{H_0}{H_{\phantom{0}}}\right)^2\left(1-\Omega_m-\frac12\frac{\Omega_m}{a^3}\right),\\
j&=&\left(\frac{H_0}{H_{\phantom{0}}}\right)^3\left(1-\Omega_m+\frac{\Omega_m}{a^3}\right)^{\frac32},\\
s&=&\left(\frac{H_0}{H_{\phantom{0}}}\right)^4\left(1-2\Omega_m-\frac52\frac{\Omega_m}{a^3}+\Omega_m^2\right.\nonumber\\
&\,&\,\,\,\,\,\,\,\,\qquad\left.+\frac52\frac{\Omega_m^2}{a^3}-\frac72\frac{\Omega_m^2}{a^6}\right),\\
l&=&\left(\frac{H_0}{H_{\phantom{0}}}\right)^5\left(1-2\Omega_m+5\frac{\Omega_m}{a^3}+\Omega_m^2\right.\nonumber\\
&\,&\,\,\,\,\,\,\,\,\qquad\left.-5\frac{\Omega_m^2}{a^3}+\frac{35}{2}\frac{\Omega_m^2}{a^6}\right)\nonumber\\
&\,&\,\,\,\,\,\,\,\,\qquad\times\sqrt{1-\Omega_m+\frac{\Omega_m}{a^3}},
\end{eqnarray}
which, evaluated at the present time, give
\cite{Capozziello:2008qc}
\begin{eqnarray}
q_0&=&-1+\frac32\Omega_m,\\
j_0&=&1,\\
s_0&=&1-\frac92\Omega_m,\\
l_0&=&1+3\Omega_m+\frac{27}{2}\Omega_m^2.
\end{eqnarray}

Inserting the previous equations in the equations (3.22)-(3.31),
we obtain
\begin{eqnarray}
\mathcal{A}_0&=&-\frac{63}{4}\Omega_m^2-\frac{27}{8}\Omega_m^3-\frac{243}{16}\Omega_m^4,\\
\mathcal{B}_0&=&-63\Omega_m^2+\frac{27}{2}\Omega_m^3+\frac{81}{16}\Omega_m^4+\frac{729}{32}\Omega_m^5,\\
\mathcal{C}_0&=&63\Omega_m^2+\frac{81}{8}\Omega_m^3-\frac{81}{16}\Omega_m^4,\\
\mathcal{A}_2&=& \frac{63}{2}\Omega_m+\frac{81}{4}\Omega_m^2,\\
\mathcal{B}_2&=&-\frac{63}{2}\Omega_m^2-\frac{81}{4}\Omega_m^3,\\
\mathcal{C}_2&=&189\Omega_m-\frac{441}{4}\Omega_m^2-84-\frac{27}{2}\Omega_m^3,\\
\mathcal{A}_3&=&\frac{243}{4}\Omega_m^2,\\
\mathcal{B}_3&=&-\frac{243}{4}\Omega_m^3,\\
\mathcal{C}_3&=&-\frac{351}{8}\Omega_m^3-\frac{81}{16}\Omega_m^4,\\
\mathcal{D}&=&63\Omega_m^2+\frac{135}{4}\Omega_m^3+\frac{243}{16}\Omega_m^4.
\end{eqnarray}
It can be checked that if $1\ll r_c$; i.e. in absence of an extra
dimension,   the function $f(R)$ reduces to $f(R)\sim R-2\Lambda$
because $f''(R_0)=0$ and $f'''(R_0)=0$. This can be assumed as a
consistency check.  However, as soon as the effect of the extra
dimension is switched on, i.e. $r_c$ is finite, the coefficients
$\mathcal{C}_i$ with $i=0,2,3$ play a crucial in defining the
shape of the function $f(R)$. Indeed, we obtain
\begin{eqnarray}
F_{GR_0}&\equiv&-\frac{\mathcal{A}_0\Omega_m+\mathcal{B}_0}{\mathcal{D}} \\
&=& -\frac12\Omega_m+1= \frac{R_0-2\Lambda}{6H_0^2},\\
F_{GR_2}&\equiv&-\frac{\mathcal{A}_2\Omega_m+\mathcal{B}_2}{\mathcal{D}}=0,\\
F_{GR_3}&\equiv&-\frac{\mathcal{A}_3\Omega_m+\mathcal{B}_3}{(j_0-q_0-2)\mathcal{D}}=0,\\
F_{IG_0}&\equiv&-\frac{\mathcal{C}_0(r_cH_0)^{-1}}{\mathcal{D}}\\
&\equiv&\frac{-112-18\Omega_m+9\Omega_m^2}{112+60\Omega_m+27\Omega_m^2}(r_cH_0)^{-1},\\
F_{IG_2}&\equiv&-\frac{\mathcal{C}_2(r_cH_0)^{-1}}{\mathcal{D}}\\
&=&\frac{4}{3}\frac{-252\Omega_m+147\Omega_m^2+112+18\Omega_m^3}{\Omega_m^2(112+60\Omega_m+27\Omega_m^2)}(r_cH_0)^{-1},\nonumber \\ \\
F_{IG_3}&\equiv&-\frac{\mathcal{C}_3(r_cH_0)^{-1}}{(j_0-q_0-2)\mathcal{D}}\\
&=& 3\Omega_m\frac{26+3\Omega_m}{112+60\Omega_m+27\Omega_m^2}(r_cH_0)^{-1}.
\end{eqnarray}
For clarity, we have split the right hand side (rhs) of Eq.~(3.20)
into two pieces: $F_{GR_0}$ and $F_{IG_0}$, the first one takes
into account the pure relativistic contribution while the second
one takes into account the effect of the extra dimension. A
similar procedure has been followed with the rhs of Eqs.~(3.21)
and (3.22).

We consider the  following observational conservative values
$\Omega_m=0.266$ and $\Omega_{r_c}=10^{-4}$ where
$\Omega_{r_c}=(4r_cH_0^2)^{-1}$
\cite{Spergel:2003cb,Lazkoz:2007zk} and we obtain  the values
reported below:
\begin{eqnarray*}
F_{GR_0} &=& 0.867\,,\\
F_{GR_2} &=& 0 \,,\\
F_{GR_3} &=& 0\,, \\
F_{IG_0} &=&  -0.018\,,\\
F_{IG_2} &=&  0.161\,, \\
F_{IG_3}  &=& 0.003\,,
\end{eqnarray*}
with the errors evaluated as in
\cite{Spergel:2003cb,Lazkoz:2007zk}. The previous results show
that, although $F_{IG_i}$  are different from zero, they are
relatively small in comparison with the present day main
contribution $F_{GR_0}$; i.e. the standard relativistic term. In
summary, the model deviates just slightly from the pure
$\Lambda$DGP model\footnote{The $\Lambda$DGP model corresponds to
the normal DGP branch endowed with a cosmological constant  and
filled with matter.} \cite{Sahni:2002dx,Lue:2004za}. This small
deviation is enough to obtain  self-acceleration without invoking
any kind of dark energy contribution on the brane. On the other
hand, if a similar analysis is carried out for a given $f(R)$
function in a 4-dimensional model, it turns out that the
$f(R)$-term match completely that of a Hilbert-Einstein action
plus a cosmological constant \cite{Capozziello:2008qc}. Most
importantly, we see that the model we have analyzed is consistent
with the $\Lambda$CDM model because the cosmographic parameters of
the $\Lambda$CDM can be matched to those of an $f(R)$ brane-world
scenario.

\section{Observational constraints}

In order to constrain the model, i.e. to  estimate the function $f(R)$ through its own value and that of its derivatives at the present time, we need to constrain observationally the cosmographic parameters by using  appropriate distance indicators. Moreover, we must take care that the expansion of the distance related quantities in terms of $(q_0, j_0, s_0, l_0)$ closely follows the exact expressions over the range probed by the data used. Taking SNeIa and a fiducial $\Lambda$CDM model as a test case, one has to check that the approximated luminosity distance\footnote{See \cite{Capozziello:2008qc} for the analytical expression.} deviates from the $\Lambda$CDM one less than the measurement uncertainties up to $z \simeq  1.5$ to avoid introducing any systematic bias. Since we are interested in constraining $(q_0, j_0, s_0, l_0)$, we will expand the luminosity distance $D_L$ up to the fifth order in $z$ which indeed allows us to track the $\Lambda$CDM expression with an error less than $1\%$ over the full redshift range. We have checked that this is the case also for the angular diameter distance $D_A = D_L(z)/(1 + z)^2$ and the Hubble parameter $H(z)$ which, however, we expand only up to the fourth order to avoid introducing a further cosmographic parameter.

In order to constrain the parameters $(h, q_0, j_0, s_0, l_0)$, we use both the Union2 SNeIa dataset \cite{Union2} and the BAO data from the analysis of the SDSS seventh release \cite{P10}. We then consider the following likelihood function\,:

\begin{table}[t]
\begin{center}
\begin{tabular}{cccccc}
\hline
$x$ & $x_{BF}$ & $\langle x \rangle$ & $x_{med}$ & $68\%$ CL & $95\%$ CL \\
\hline \hline
~ & ~ & ~ & ~ & ~ & ~ \\
$h$ & 0.744 & 0.750 & 0.750 & (0.725, 0.775) & (0.701, 0.802) \\
~ & ~ & ~ & ~ & ~ & ~ \\
$q_0$ & -0.43 & -0.44 & -0.45 & (-0.48, -0.41) & (-0.51, -0.36) \\
~ & ~ & ~ & ~ & ~ & ~ \\
$j_0$ & -0.35 & 0.01 & 0.01 & (-0.11, 0.14) & (-0.33, 0.35) \\
~ & ~ & ~ & ~ & ~ & ~ \\
$s_0$ & -1.3 & 0.4 & 0.4 & (-0.3, 1.0) & (-1.2, 1.8) \\
~ & ~ & ~ & ~ & ~ & ~ \\
$l_0$ & 14.7 & -0.6 & -1.0 & (-4.6, 3.7) & (-11.3, 11.7) \\
~ & ~ & ~ & ~ & ~ & ~ \\
\hline
\end{tabular}
\end{center}
\caption{Constraints on the cosmographic parameters by jointly fitting the Union2 SNeIa sample and the BAO data. Columns are as follows\,: 1. parameter id; 2. best fit; 3., 4. mean and 
median from the marginalized likelihood; 5., 6. $68$ and $95\%$ confidence ranges.}
\end{table}

\begin{equation}
{\cal{L}}(p) = {\cal{L}}_{SNeIa}({\bf p}) \ \times \ {\cal{L}}_{BAO}({\bf p})
\label{eq: deflike}
\end{equation}
where ${\bf p}$ is the set of model parameters and we have defined the likelihood function for the probe $i$ as\,:

\begin{equation}
{\cal{L}}_{i}({\bf p}) = \frac{1}{(2 \pi)^{{\cal{N}}_{i}/2} |{\bf C}_{i}|^{1/2}}
\exp{\left ( - \frac{{\bf \Delta}_{i}^{T} {\bf C}_{i}^{-1} {\bf \Delta}_{i}}{2} \right )} \ .
\label{eq: likei}
\end{equation}
For SNeIa, ${\bf \Delta}_{SNeIa}$ is ${\cal{N}}_{SNeIa}$ (with ${\cal{N}}_{SNeIa} = 557$) column vector with elements computed as\,:

\begin{equation}
{\bf \Delta}_{SNeIa, j} = \mu_{obs}(z_j) - \mu_{th}(z_j, {\bf p})
\label{eq: defdeltasneia}
\end{equation}

\begin{equation}
\mu_{th}(z) = 25 + 5 \log{D_L(z, {\bf p})} \ ,
\label{eq: muth}
\end{equation}
while the ${\bf C}_{SNeIa}$ is a diagonal matrix. For BAO, we set\,:

\begin{equation}
{\bf \Delta}_{BAO, j} = d_{obs}(z) - d_{th}(z_j, {\bf p})
\label{eq: defdeltabao}
\end{equation}

\begin{equation}
d_{th}(z, {\bf p}) = \frac{r_s(z_d)}{D_V(z, {\bf p})} = r_s(z_d) \left [ \frac{(1 + z)^2 D_A^2(z, {\bf p}) c z}{H(z, {\bf p})} \right ]^{-1/3} \ ,
\label{eq: defdz}
\end{equation}
where we set the sound horizon distance to the drag redshift as $r_s(z_d) = 152.6 \ {\rm Mpc}$. Percival et al. \cite{P10} provide estimates of $d_z$ for $z = (0.20, 0.35)$ and the corresponding covariance matrix that we use as input in Eq.(\ref{eq: defdeltabao}). We remember the reader that we use a fifth order expansion in $z$ for both $D_L(z)$ and $D_A(z)$, while $H(z)$ is expanded to the fourth order only. Since the BAO data are at low redshift, the resulting approximated expression for $d_{th}(z)$ closely follows the exact values. Finally, we also use a Gaussian prior on $h$ from local distance measurement so that (\ref{eq: likei}) reduces to a Gaussian centred on $h = 0.742$ and with variance $\sigma_h = 0.036$ \cite{shoes}.

In order to sample the five dimensional parameter space, we use a Markov Chain Monte Carlo algorithm running two chains (with 125000 point each) and checking the convergence according to the Gelman\,-\,Rubin criterium ($R - 1 < 0.1$). The resulting constraints are summarized in Table I where we give the best fit parameters and the constraints over the single $p_i$ obtained by marginalizing over the other ones. As a general remark, we find that these constraints are in agreement with previous constraints in literature \cite{Vitagliano:2009et,XW10}. Note, however, that our confidence ranges turn out to be narrower than usually found. This is likely due to our inclusion of the lerk parameter $l_0$. In a sense, we are now better approximating the (unknown) actual distances and Hubble parameter so that not all the possible combinations of $(h, q_0, j_0, s_0)$ are possible, but only the ones that are compatible with the constrained $l_0$.

\begin{figure*}
\centering
\subfigure{\includegraphics[width=5cm]{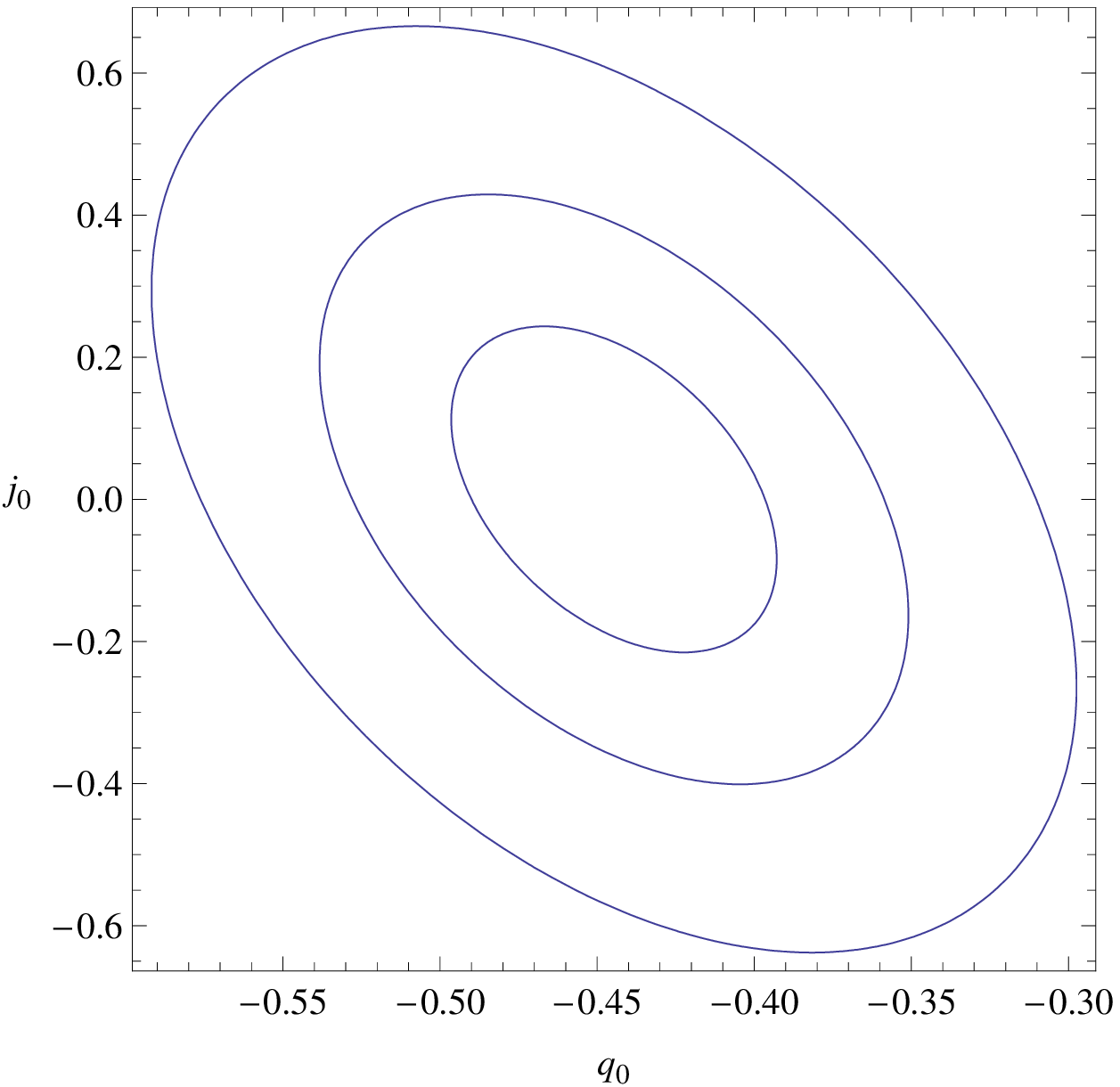}} \goodgap
\subfigure{\includegraphics[width=5cm]{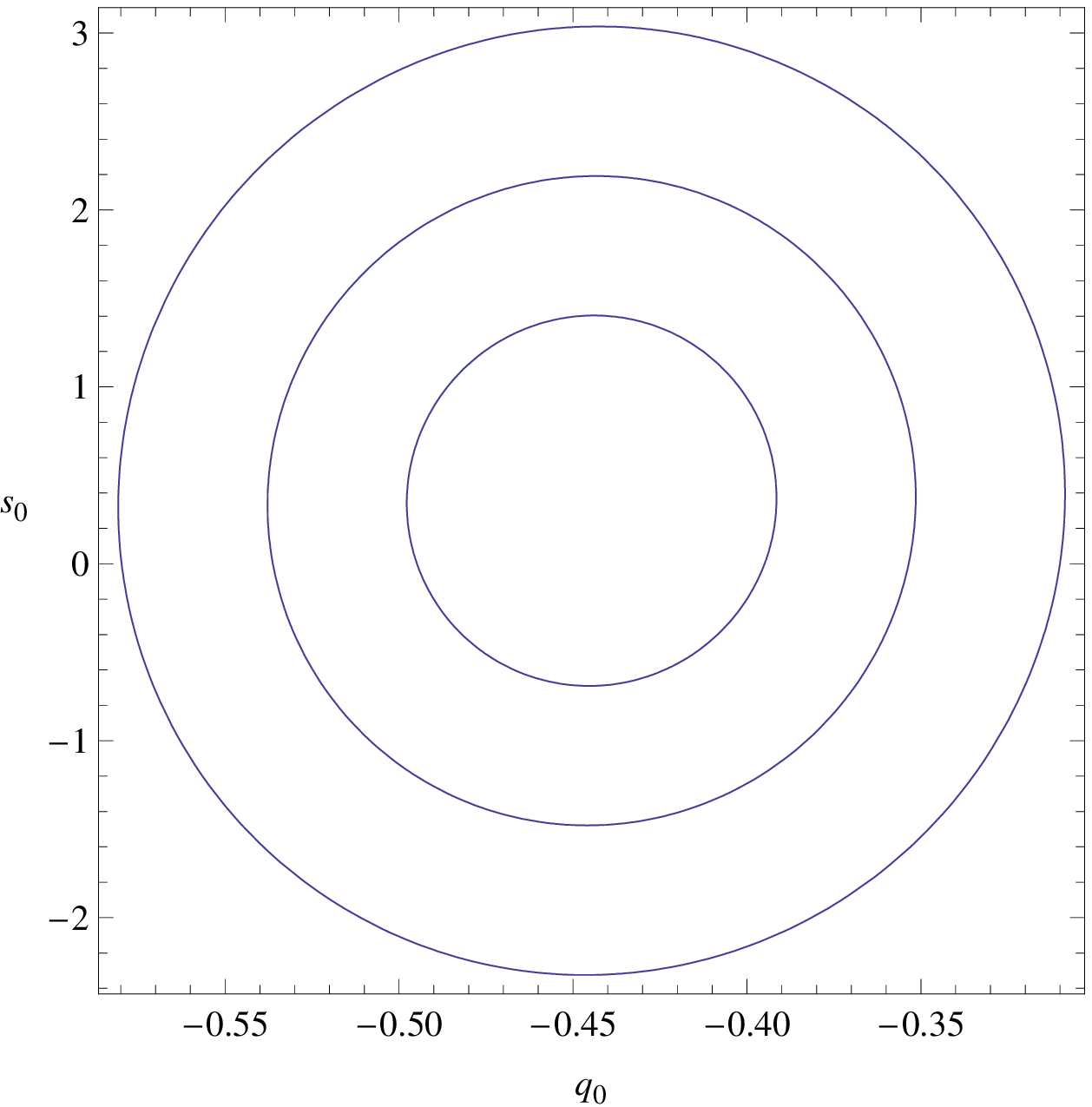}} \goodgap
\subfigure{\includegraphics[width=5cm]{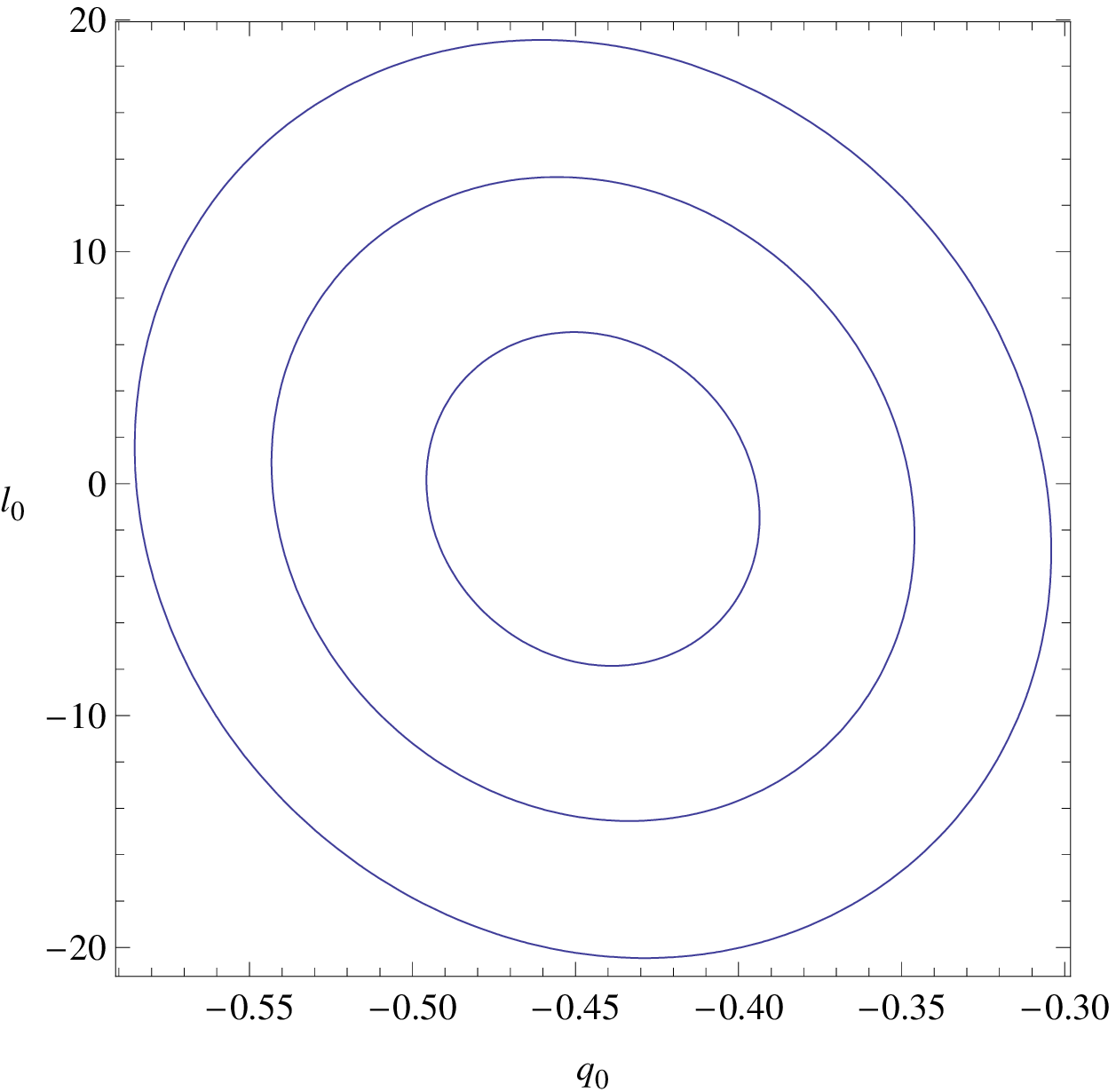}} \goodgap \\
\caption{Isolikelihood ($68$, $95$ and $99\%$ CL) contours for the fit to the SNeIa and BAO data. In each panel, we marginalize  the given cosmographic parameter with respect to $q_0$.}
\label{fig: mcplots}
\end{figure*}

In order to translate our constraints on the cosmographic parameters on similar constraints on $f(R)$ and its derivatives, we should just use Eqs.({\ref{eq: frfinal})\,-\,(\ref{eq: a3b3c3}) evaluating them along the final coadded and thinned chain and then looking at the corresponding histograms. To this end, however, we should set also the values of $\Omega_M$ and $\Omega_{r_c}$ (and hence $r_c H_0 = 1/2 \sqrt{\Omega_{r_c}}$) which are not constrained by the fitting analysis described before. To partially overcome this difficulty, we adopt the following strategy. Defining for shortness

\begin{displaymath}
f_0 = \frac{f(R_0)}{6 H_0^2} \ \ , \ \ f_2 = \frac{f^{\prime \prime}(R_0)}{(6 H_0^2)^{-1}} \ \ , \ \
f_3 = \frac{f^{\prime \prime \prime}(R_0)}{(6 H_0^2)^{-2}} \ \ ,
\end{displaymath}
we first constrain these quantities setting $\Omega_{r_c} = 10^{-4}$ and varying $\Omega_M$ along the chain using $\Omega_M = \omega_M h^{-2}$ with the physical matter density $\omega_M = 0.1329$ in agreement with the WMAP7 data. Note that we are neglecting the uncertainty on $\omega_M$ since it is much lower than those on the cosmographic parameters. We also stress that, although the fiducial value for $\omega_M$ has been obtained for a $\Lambda$CDM model, it should be unchanged for any model which reduces to the GR\,+\,matter domination at the CMBR epoch as is our case. We can then scale the results to a different value of $r_c H_0$ noting that, by simple algebra, we get from Eq.(\ref{eq: frfinal})\,:

\begin{equation}
\frac{f_i}{f_{i}^{fid}} = \alpha_i \frac{(r_c H_0)_{fid}}{r_c H_0} + \beta_i = \alpha_i \left ( \frac{\Omega_{r_c}^{fid}}{\Omega_{r_c}}
\right )^{1/2} + \beta_i
\label{eq: scaling}
\end{equation}
with the quantities labelled $fid$ are obtained for the fiducial $\Omega_{r_c}$ value and we have defined (for $i = 0, 2, 3$)\,:

\begin{table}[t]
\begin{center}
\begin{tabular}{cccccc}
\hline
$x$ & $x_{BF}$ & $\langle x \rangle$ & $x_{med}$ & $68\%$ CL & $95\%$ CL \\
\hline \hline
~ & ~ & ~ & ~ & ~ & ~ \\
$f_{0}$ & 0.897 & 0.912 & 0.912 & (0.876, 0.949) & (0.828, 0.992) \\
~ & ~ & ~ & ~ & ~ & ~ \\
$f_{2}$ & 0.126 & 0.163 & 0.161 & (0.140, 0.185) & (0.116, 0.220) \\
~ & ~ & ~ & ~ & ~ & ~ \\
$f_{3}$ & -0.130 & -0.139 & -0.142 & (-0.181, -0.101) & (-0.240, -0.004) \\
~ & ~ & ~ & ~ & ~ & ~ \\
$\alpha_0$ & -0.0190 & -0.0168 & -0.0167 & (-0.0180, -0.0155) & (-0.0201, -0.0141) \\
~ & ~ & ~ & ~ & ~ & ~ \\
$\beta_0$ & 1.0190 & 1.0168 & 1.0167 & (1.0155, 1.0180) & (1.0141, 1.0201) \\
~ & ~ & ~ & ~ & ~ & ~ \\
$\alpha_2$ & 0.0130 & 0.0190 & 0.0191 & (0.0172, 0.0209) & (0.0137, 0.0236) \\
~ & ~ & ~ & ~ & ~ & ~ \\
$\beta_2$ & 0.9870 & 0.9810 & 0.9809 & (0.9791, 0.9828) & (0.9764, 0.9863) \\
~ & ~ & ~ & ~ & ~ & ~ \\
$\alpha_3$ & 3.1091 & 0.0071 & 0.0100 & (0.0037, 0.0140) & (-0.0272, 0.0263) \\
~ & ~ & ~ & ~ & ~ & ~ \\
$\beta_3$ & -2.1090 & 0.9929 & 0.9899 & (0.9860, 0.9962) & (0.9736, 1.0272) \\
~ & ~ & ~ & ~ & ~ & ~ \\
\hline
\end{tabular}
\end{center}
\caption{Constraints on the fiducial $f_i$ values and on the scaling coefficients $(\alpha_i, \beta_i)$ from the Markov Chain for the cosmographic parameters. Columns are as in Table I.}
\end{table}

\begin{equation}
\alpha_i = \frac{{\cal{C}}_i}{\left ( {\cal{A}}_i \Omega_M + {\cal{B}}_i \right ) (r_c H_0)_{fid} + {\cal{C}}_i}
\label{eq: defalpha}
\end{equation}

\begin{equation}
\beta_i = \frac{\left ( {\cal{A}}_i \Omega_M + {\cal{B}}_i \right ) (r_c H_0)_{fid}}{\left ( {\cal{A}}_i \Omega_M + {\cal{B}}_i \right ) (r_c H_0)_{fid} + {\cal{C}}_i} \ .
\label{eq: defbeta}
\end{equation}
The constraints on the fiducial $f_i$ and the scaling parameters $(\alpha_i, \beta_i)$ obtained by evaluating these quantities along the Markov chain for the cosmographic parameters are summarized in Table II. Considering the median values and the quite narrow confidence ranges, we find that that the scaling parameters $(\alpha_i, \beta_i)$ are well consistent with the $f_i$ being linear functions of the inverse of the crossover scale $r_c$  hence allowing us to easily estimate the impact of uncertainties on this parameter on the final estimate of the present day values of $f(R)$ and its derivatives. Somewhat surprisingly, the fiducial $f_i$ are reasonably well constrained notwithstanding the large uncertainties on the cosmographic parameters. Such a result can be qualitatively understood noting that $f_i$ depend on $(q_0, j_0, s_0, l_0)$ through a ratio of coefficients so that it is possible that a variation in the numerator is compensated by a similar variation in the denominator in such a way that the final $f_i$ is unaltered. As a consequence, the dependence on the cosmographic parameters is made weaker thus reducing the impact of the parameters uncertainties. \\

\section{Conclusions}

Cosmography is a useful method to give a picture of the observed
universe considering minimal assumptions (isotropy, homogeneity,
Taylor series expansion of distances) without choosing any
dynamical model a priori.

In this paper, we have taken into account the problem to test
brane-cosmology, where an $f(R)$-term is present in the boundary
4D-action, by cosmography. Being $\Lambda$CDM a realistic picture
of the today observed universe, we have adopted $\Lambda$CDM
observational results as priors for our approach. We assumed the
$f(R)$ function to be analytical in order to evaluate the
higher-order curvature contributions with respect to general
relativity contribution, i.e. $f(R)= R$. The results are
encouraging since small higher-order deviations with respect to
general relativity give dynamical behaviors, consistent with
observed cosmic acceleration, {\it without}  introducing  dark
energy terms.

However, the approach should be consistently probed at small,
medium and high redshift by selecting suitable standard candles
or, at least, reliable distance indicators at any scale. Despite
of this technical difficulty, the method outlined here  deserves
further investigations since it is connecting a fundamental
theory, as the DGP-brane model,  with data coming from precision
cosmology. We have here addressed this point in a preliminary way
by only using SNeIa and BAO, but other probes (such as GRBs) may be added to further
narrow the constraints on the present day values of $f(R)$ and its second and
third derivatives with respect to $R$.

\section*{Acknowledgments}

MBL is  supported by the Portuguese Agency Funda\c{c}\~{a}o para a
Ci\^{e}ncia e Tecnologia through the fellowship
SFRH/BPD/26542/2006. She also wishes to acknowledge the
hospitality of LeCosPA at the National University of Taiwan during
the completion of part of this work. SC acknowledges INFN support
for this research.

\end{document}